\documentclass[referee]{raa}
\usepackage{graphicx,times}
\usepackage{natbib}
\usepackage{amssymb,amsmath}
\usepackage{ulem}
\usepackage{color}

\begin{document}

    \title{Irregular changes  in H$\alpha$ emission line of V423 Aur observed by LAMOST Medium-Resolution 
    Spectrographs}

 \volnopage{ {\bf 2012} Vol.\ {\bf X} No. {\bf XX}, 000--000}
   \setcounter{page}{1}

\author{Chao-Jian Wu\inst{1,2}, Hong Wu\inst{1,2}, 
Chih-Hao Hsia\inst{3}, Wei Zhang\inst{1,2}, Juan-Juan Ren\inst{1,2}, Guang-Wei Li\inst{1,2}, 
Jian-Jun, Chen\inst{1,2}, Fan, Yang\inst{1,2}, Jian-Rong, Shi\inst{1,2}, 
Yong-Hui Hou\inst{4,5}, Ji-Feng Liu\inst{1,2,5}}

\institute{Key Laboratory of Optical Astronomy, 
National Astronomical Observatories, Chinese Academy of Sciences，
Beijing 100101, China; {\it chjwu@bao.ac.cn}\\
\and
National Astronomical Observatories, 
Chinese Academy of Sciences, Beijing 100012, China\\
\and
Space Science Institute, Macau University of Science and Technology, Taipa, Macau\\
\and
    Nanjing Institute of Astronomical Optics, \& Technology, National Astronomical Observatories, 
    Chinese Academy of Sciences, Nanjing 210042, China\\
    \and
   School of Astronomy and Space Science, University of Chinese Academy of Sciences\\
\vs \no
{\small Received ---; accepted ---}
}

\abstract{
    We obtained 7 spectra  of the Be star V423 Aur on Dec. 5th, 2017 using the LAMOST 
    Medium-Resolution Spectrograph with exposures from 600 to 1200 seconds. 
    These spectra show the irregular H$\alpha$ emission line profile variations (LPVs). 
    In the seven spectra, from the 4th to 7th, the left part of H$\alpha$ profile even show 
    excess. However, no variation can be seen from the follow-up observation of photometry by 
    1.26-m telescope and High-Resolution spectra by 2.16-m telescope. According to the 
    High-Resolution spectra, we conclude that it is a B7V type star with E(B-V)= 0.709$\pm$0.036 
    and its $v\mathrm{sin}i$ is $\sim$ 221.8km/s. The short-term H$\alpha$ LPVs  
    could be explained as a result of the transient ejection of matter from rotating 
    disk or shell around V423 Aur.
\keywords{stars: emission-line, Be --- stars: variables: general --- stars: rotation}
}
\authorrunning{Ch.-J. Wu et al.}
\titlerunning{ V423 Aur observed by LAMOST}
\maketitle

\section{Introduction}
\label{s:intro}

The stellar variability 
can be used to study a variety of physical processes. 
Stellar variability mainly due to several reasons: 
asteroseismology , eruptive behavior, the eclipsing binaries 
and period-luminosity relations of evolved stars, the 
explosive behavior of novae and supernovae, and so on 
\citep{2018conroy, 2008eyer, 2015catelan}. The lines variations, 
which are from circumstellar emission and absorption lines, 
are also very common to Be stars, precisely speaking, the 
classical Be stars \citep{2003Porter}. 

The classical Be stars, are a 
heterogeneous set of non-supergiant stars with B spectral 
types and one or more Balmer emission lines \citep{1981Jaschek}.
Besides B type and emission lines, rotation, variation and 
circumstellar gas are also the main features of Be stars. 

Rotation is an important feature of Be star \citep{1931Struve}, 
which may be a 
main contributor to the generation of the circumstellar medium.
But whether or not Be stars rotate at the 
critical limit (the gravity balances centrifugal force) is still not clear 
\citep{2003Porter}. Some results of studies show that  Be stars 
rotate at values of 70\% $\sim$ 80\% of their critical 
rotation\citep{}. Though the rotation is not 100\% of their 
critical rotation, it is still very high and is an 
important index to distinguish Be stars from normal B stars.

The timescale of line variation of Be stars ranges from 
minites to a few days \citep{2003Porter}. Early-type Be 
stars (earlier than B5) usually show short-term 
variation \citep{1974Lucy, 1975Balona}. 
\citet{1989Cuypers, 1992Balona} confirmed this trend.
The short-periodic line variation is considered to be 
due to nonradial pulsation in Baade's work \citep{1982Baade}.
While \citet{1990Balona, 1995Balona} attributed the variation 
to stellar spots and coratating clouds. In the work of 
\citet{2001Smith}, the results seemed to be more supportive 
of the pulsation.

In this paper, a Be star 
was observed by LAMOST medium resolution spectrographs. The medium 
resolution spectra show obviously intensity and profile changes in H$\alpha$ emission 
line (Section \ref{s:obs:mrs}). However, no changes can be found from 
the follow-up observation of photometry and  
\ref{s:data:photometry}) and High-Resolution spectra (HRS) 
(Section \ref{s:data:hrs}). Some results were concluded based on these 
observation data in Section \ref{s:results}. We also discussed the reasons of 
H$\alpha$ changes of the Be star in Section \ref{s:dis}. Summary was 
presented in Section \ref{s:summary}.

\section{Observation}
\label{s:obs}
\subsection{Medium Resolution Spectra of V423 Aur}
\label{s:obs:mrs}

LAMOST (Large Sky Area Multi-Object fiber Spectroscopic 
Telescope; also known as Guo Shou Jing telescope) is the 
first large astronomical 
scientific device in China. It has a five-degree field of view and 
can simultaneously observe the spectra of 4,000 objects. From 2011 
to 2017 (the pilot and the first 5 years regular survey), LAMOST 
has obtained more than 9 million spectra (R $\sim$ 1800). Since 
Sep. 2017, it started test observation with medium resolution 
(R $\sim$ 7500) spectrographs. There are all 16 medium resolution 
spectrographs, each spectrograph has two cameras, red and blue.  
Red camera covers the wavelength range from 4950 $\mathrm{\AA}$ to 5350 
$\mathrm{\AA}$, and blue camera covers from 6300 $\mathrm{\AA}$ to 6800 
$\mathrm{\AA}$. Since Sep. 2018, it starts the second stage survey 
(the Medium-Resolution spectral Survey,  MRS) program. 

\begin{figure*}                         
    \centering
    \includegraphics[width=0.6\textwidth]{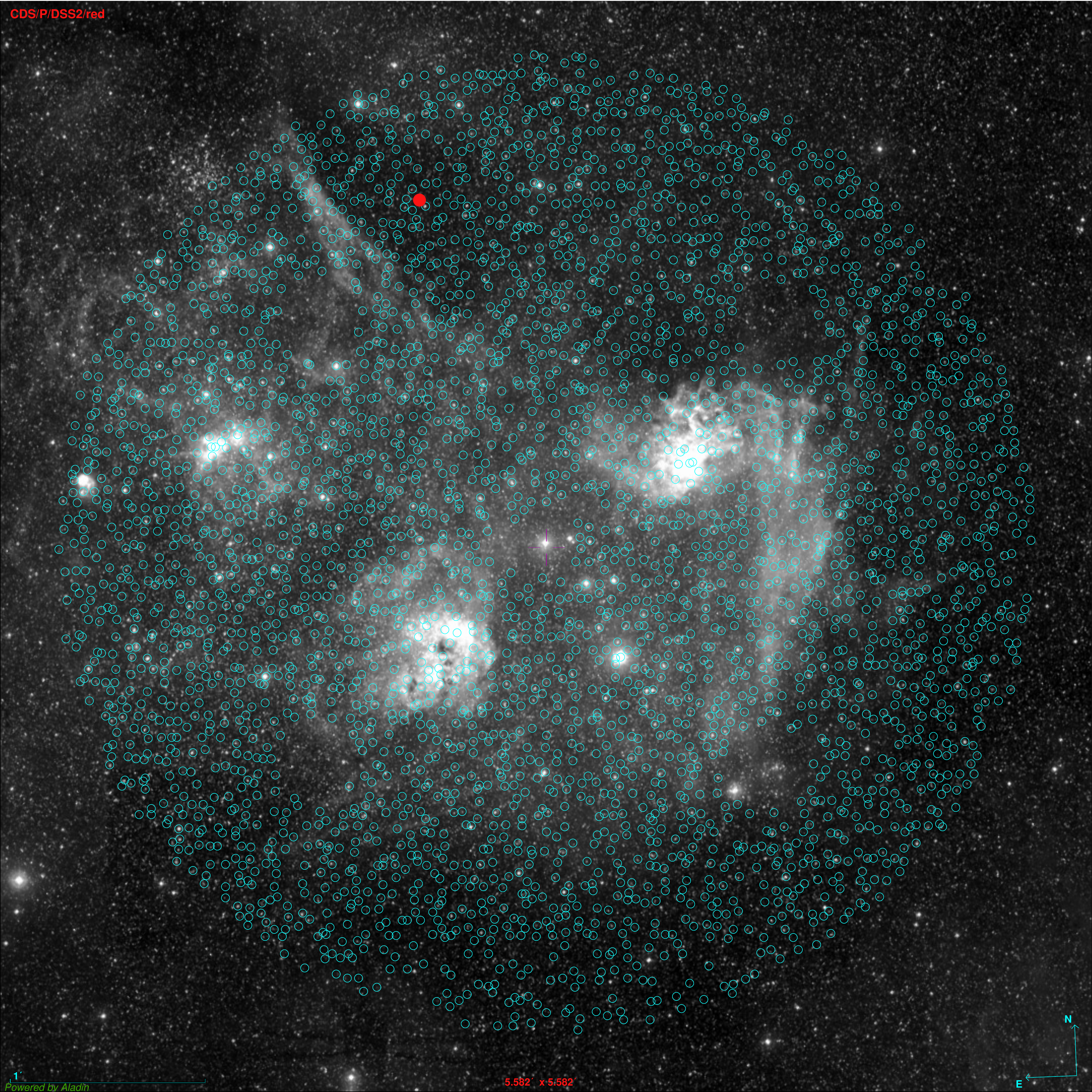}
    \caption{Fiber distribution of test sky area. Blue circles 
    are coordinates of 4000 fibers. Purple plus symbol is the center 
    star HIP24879. Red dot is V423 Aur. 
    \label{fig:lamost_plate}}
\end{figure*}

On Dec. 5th, 2017, we observed a test sky area  
with medium-resolution spectrographs of LAMOST. The 
central star of the test plate is HIP24879. 
According to the corresponding H$\alpha$ photometric image of IPHAS 
(The INT Photometric Halpha Survey of the Northern Galactic Plane), 
some bright stars and many bright positions of nebulae are selected as 
the input catalogue. The catalogue includes 47456 positions in all. 
After redistribution with LAMOST Survey Strategy System (SSS), 3478 source 
locations (1679 stars and 1799 nebula positions) from 47456 are allocated 
to 3478 fibers. The other 522 fibers are immovable. Figure 
\ref{fig:lamost_plate} shows the fiber distribution.

\begin{table}
    \centering
\caption{Observation of V423 Aur with LAMOST}
\label{tab:1}      
\begin{tabular}{ccccc}
\hline\noalign{\smallskip}
    Spectrum ID & Observation Time (UT) & MJD & Exposure    \\
\noalign{\smallskip}\hline\noalign{\smallskip}
  No.1 &  2017-12-05T15:01:19.64 & 58092.6259 & 600s  \\
  No.2 &  2017-12-05T15:14:41.29 & 58092.6352 & 600s  \\
  No.3 &  2017-12-05T15:28:03.88 & 58092.6444 & 600s  \\
  No.4 &  2017-12-05T15:47:30.74 &  58092.6580 & 900s  \\
  No.5 &  2017-12-05T16:05:52.25 & 58092.6707 & 900s  \\
  No.6 &  2017-12-05T16:24:13.89 & 58092.6835 & 900s  \\
  No.7 &  2017-12-05T16:47:35.57 & 58092.6997 & 1200s  \\

\noalign{\smallskip}\hline
\end{tabular}
\end{table}

This test sky area was observed for seven times. More detailed observation information can 
be seen in Table \ref{tab:1}. The 
exposure time of first three spectra is 600s, the second three spectra  
is 900s and the last is 1200s. In this observation, we obtained 1679 medium 
resolution stellar spectra in total. The 1D medium resolution spectra were 
processed by LAMOST pipeline. 
The procedures include standard preprocessing (like bias subtraction, 
flat-field correction, fiber efficiency correction etc), spectra extraction, 
wavelength calibration and skylight subtraction. Because the nebula is mixed 
with the skylight background, the skylight subtraction is done only for stellar 
spectra. 

\begin{figure*}                                                              
    \centering
    \includegraphics[width=\textwidth]{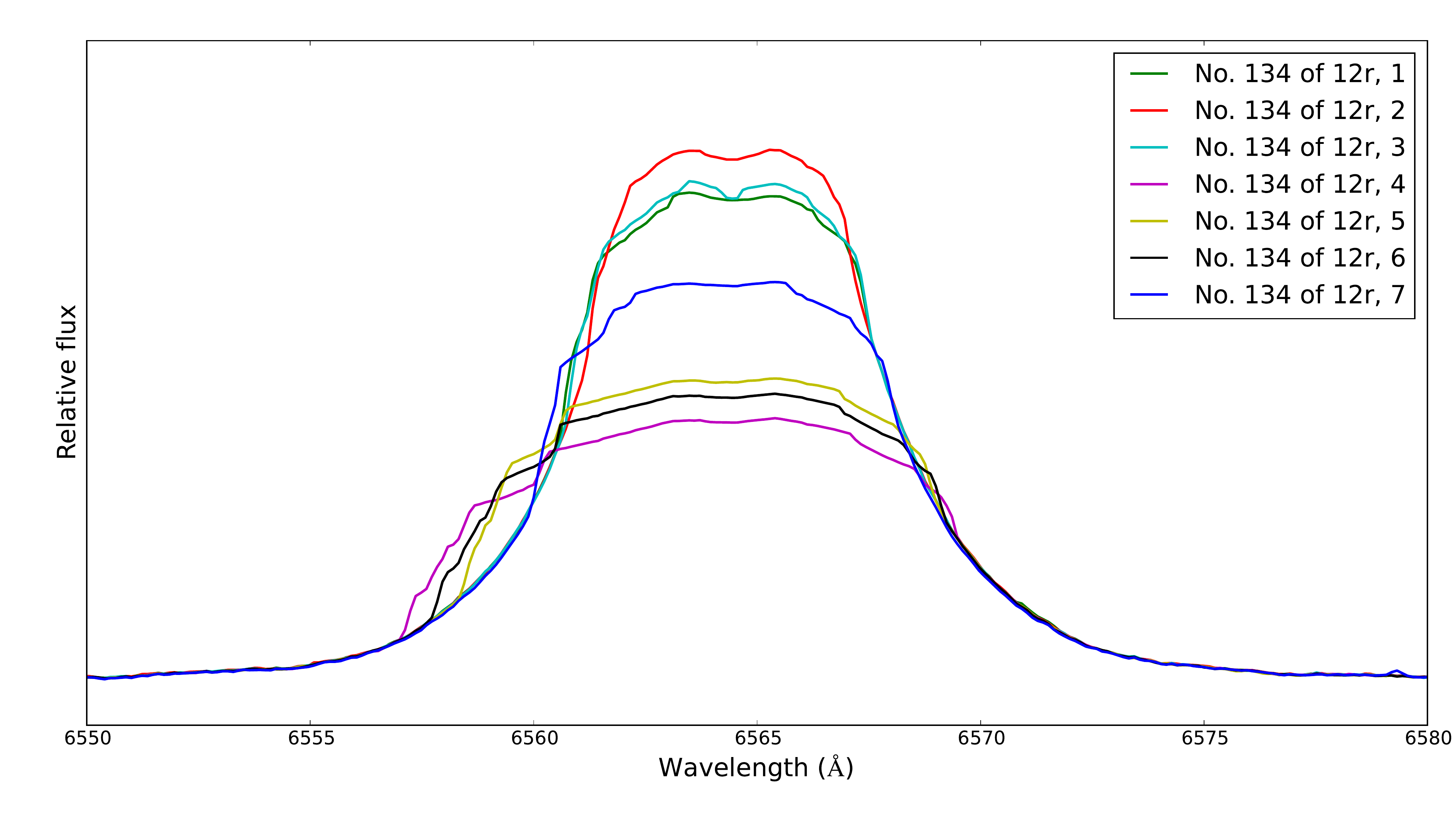}
    \caption{Variable H$\alpha$ emission line of V423 Aur observed by LAMOST .  
    The 4th (pink), 5th (yellow), 6th (black) and 7th (blue) spectrum even show broader 
    spectral line width.
    \label{fig:Ha}}
\end{figure*}

After checked all the 1D spectra of the 1679 stars, we found 
a star showing a variable H$\alpha$ emission line, see 
Figure \ref{fig:Ha}. By matching with Simbad, the star is V423 Aur 
(Red dot in Figure \ref{fig:lamost_plate}). 
As shown in Figure 
\ref{fig:Ha}, the spectra are all background corrected and the 
continuous spectra have been scaled to equal. Obviously, the H$\alpha$ 
emission lines changed irregularly. The 2nd (red) H$\alpha$ flux is highest and 
the 4th (pink) flux is lowest. The 4th (pink), 5th (yellow), 
6th (black) and 7th (blue) spectrum even show 
broader spectral line width. 

\begin{figure*}                                                              
    \centering
    \includegraphics[width=\textwidth]{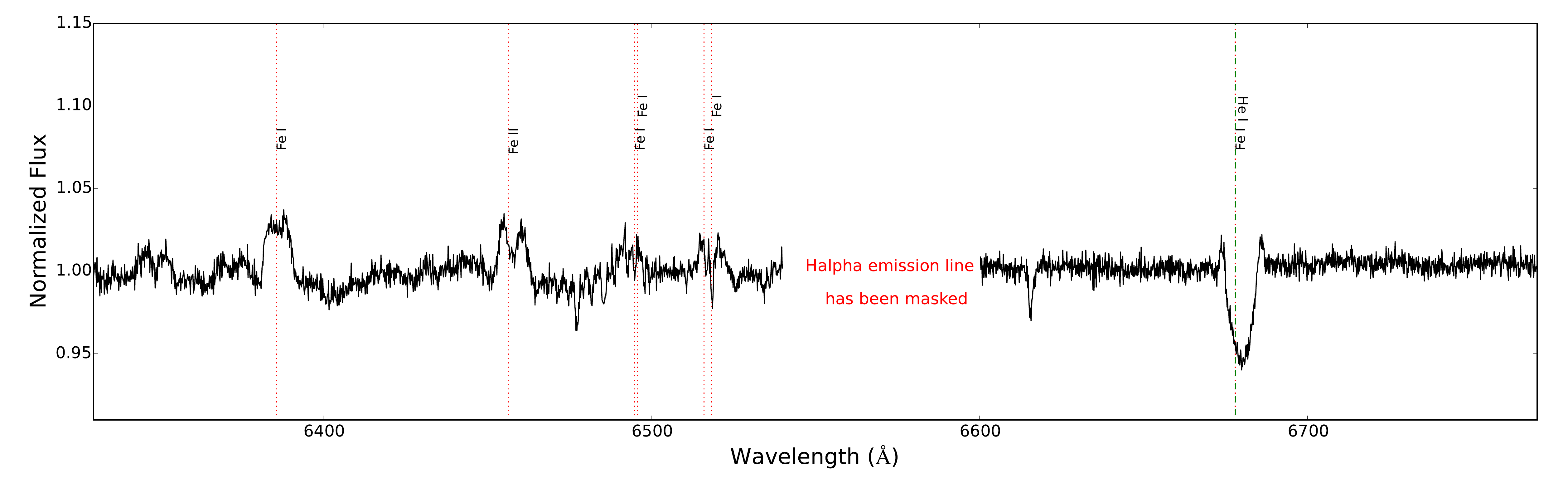}
    \caption{Medium-resolution spectrum of V423 Aur by red camera. 
    All Fe lines show bimodal structure.  The HeI absorption line centered 
    on 6678.15$\mathrm{\AA}$ was broadened. 
    \label{fig:red}}
\end{figure*}

\begin{figure*}                                                              
    \centering
    \includegraphics[width=\textwidth]{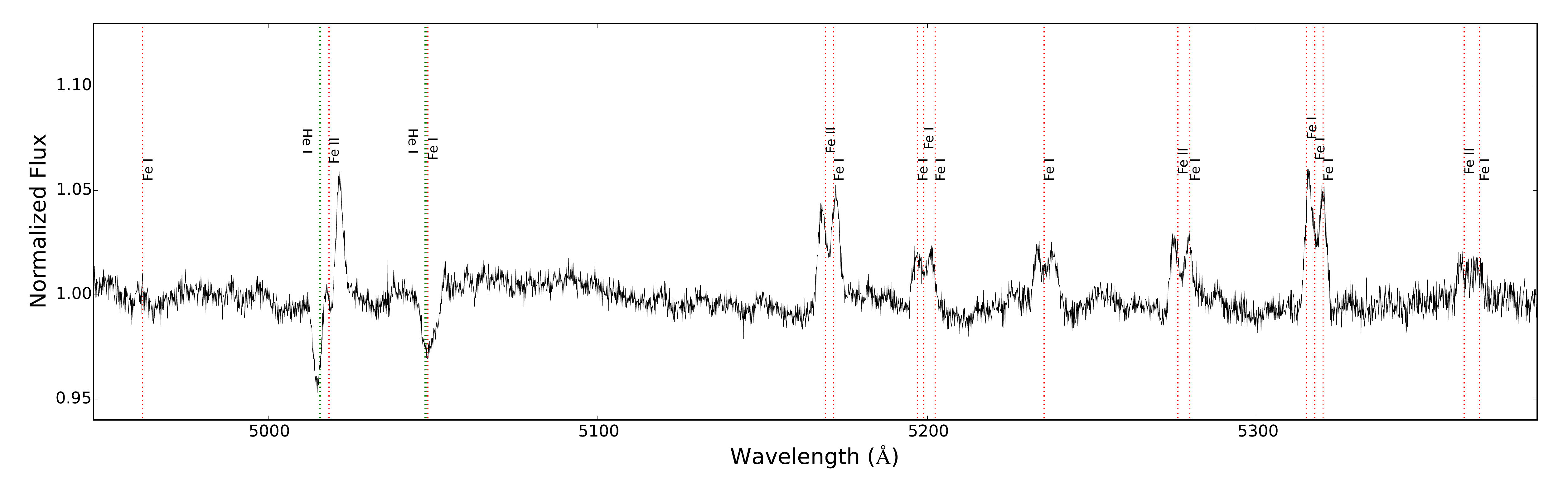}
    \caption{Medium-resolution spectrum of V423 Aur by blue camera.  
    \label{fig:blue}}
\end{figure*}

Figure \ref{fig:red} and \ref{fig:blue} show two normalized 
medium-resolution spectra of red camera and blue camera. Many 
emission lines and two strong absorption lines have been marked. 
They are almost all Fe lines and show bimodal structure. From 
the spectra, this star can be identified as a Be star.  

\subsection{Photometry of V423 Aur}
\label{s:data:photometry}

We also observed V423 Aur with 1.26-m National Astronomical Observatory-Guangzhou 
University Infrared/Optical Telescope (NAGIOT) located at XingLong 
observation, National Astronomical Observatories, Chinese Academy of 
Sciences, on Jan. 09 2018 and Jan. 11 2018.  
NAGIOT has two channels, optical and near infrared channels. The optical 
channel is splitted into three optical passbands by the TRIPOL5 instrument. 
The TRIPOL5 use three SBIG STT-8300M cameras with a CCD of 3326$\times$2504 pixels 
and a field of view of $6.0' \times 4.5'$. So NAGIOT enables simultaneous photometry 
in three optical bands (SDSS $g$, $r$ and $i$). Three images with 3 different 
filters can be obtained simultaneously. 

\begin{figure*}
    \centering
    \includegraphics[width=\textwidth]{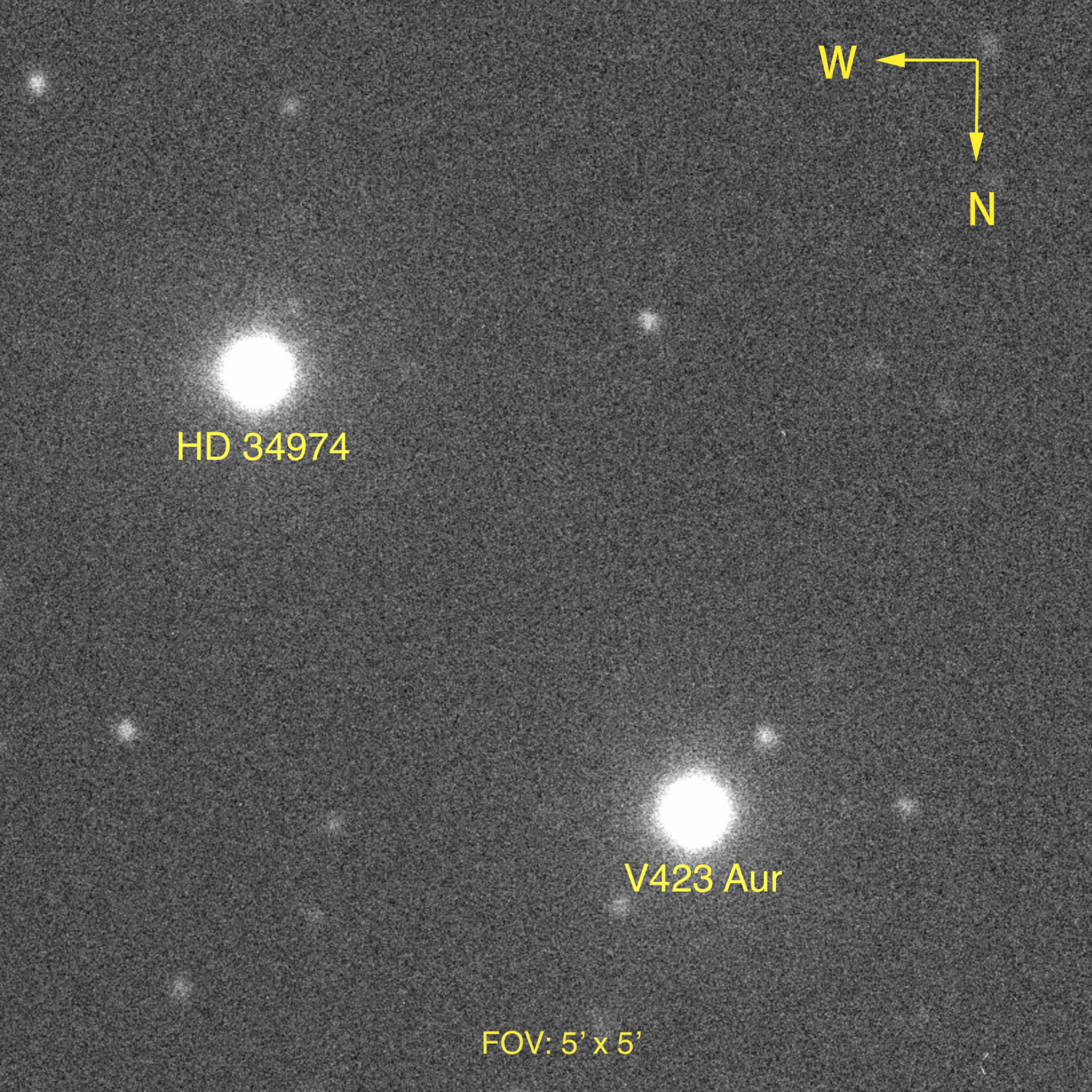}
    \caption{$r$ band image observed on 2018-01-09 with NAGIOT. 
    Two brightest stars are V423 Aur and HD34974. 
    \label{fig:r_v423}}
\end{figure*}

\begin{figure*}
    \centering
    \includegraphics[width=0.8\textwidth]{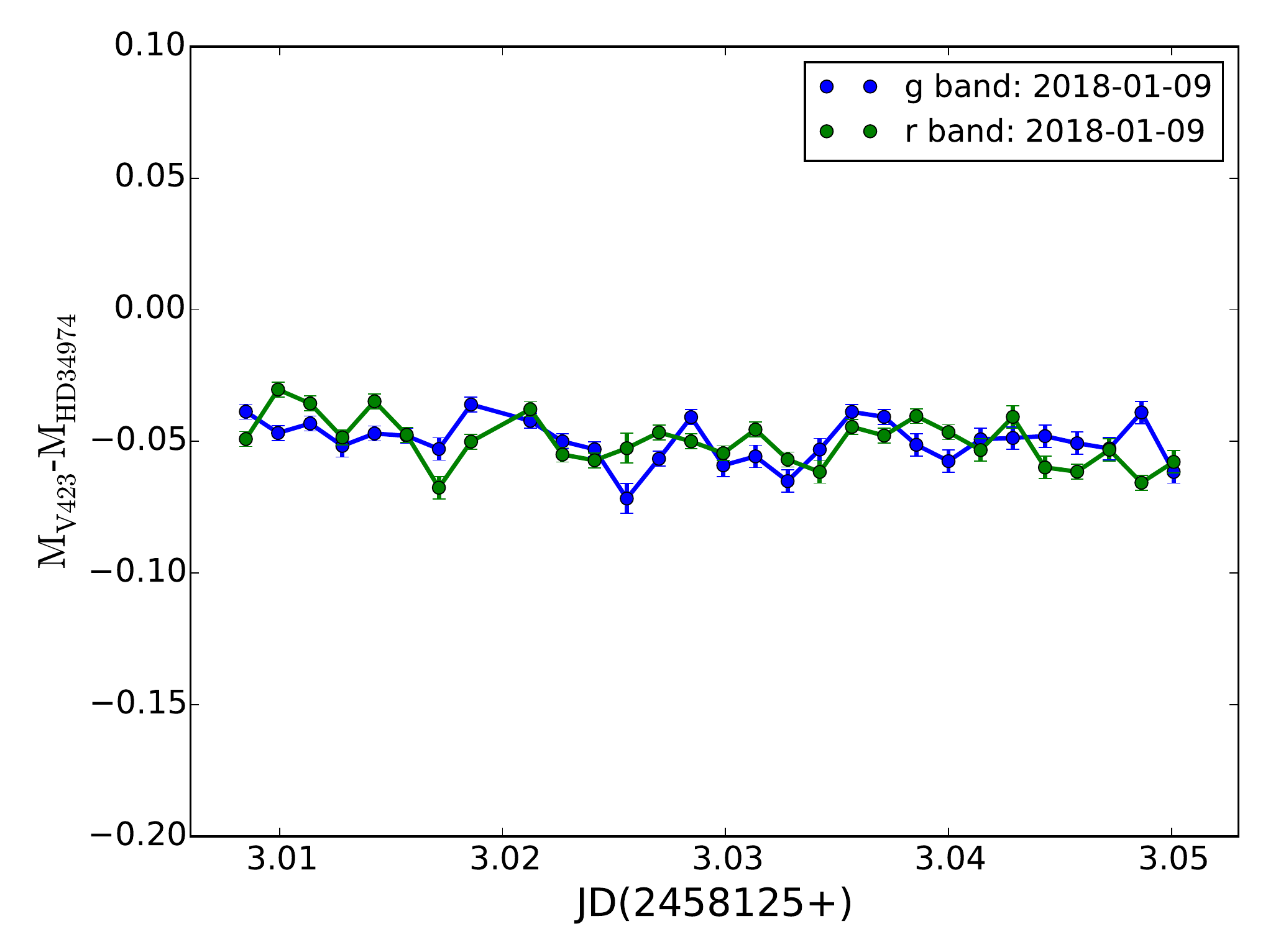}
    \caption{ $g$ and $r$ band light curves of V423 Aur observed on 2018-01-09. 
    \label{fig:09gri}}
\end{figure*}

\begin{figure*}                                     
    \centering
    \includegraphics[width=0.8\textwidth]{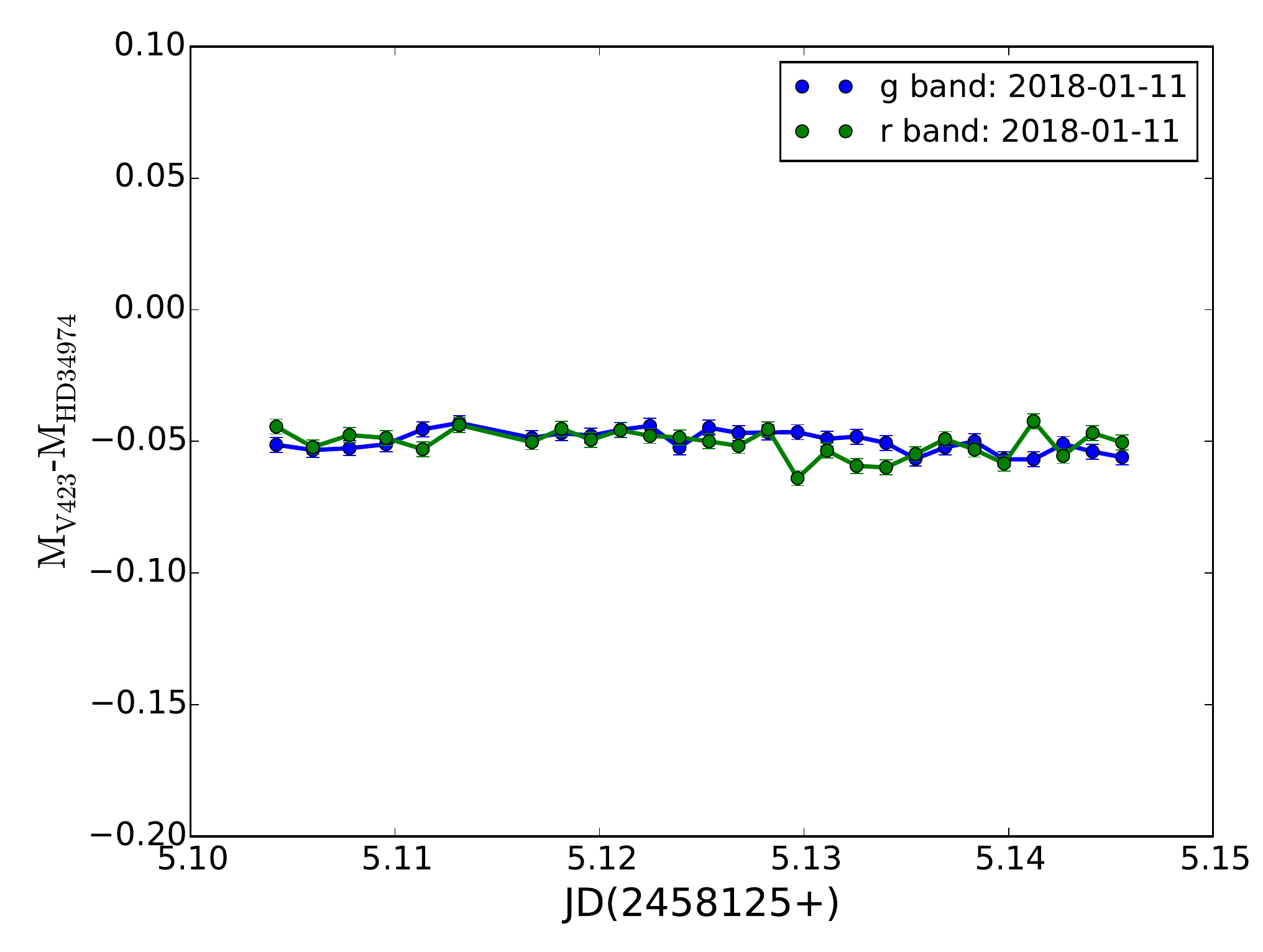}
    \caption{$g$ and $r$ band light curves of V423 Aur observed on 2018-01-11.
    \label{fig:11gri}}
\end{figure*}

V423 Aur was observed for 29 times on Jan. 09 
and 27 times on Jan. 11. Table \ref{tab:2} lists 
detailed information of observation. The data reductions were carried out 
with the PyRAF following the procedures: bias subtraction, flat-field correction. 
We need to use the results of differential photometry, so the flux calibration of 
the images was not done.
Figure \ref{fig:r_v423} (r band ) is 
one of the optical images observed by NAGIOT on Jan. 09. 
There are two brightest stars 
in Figure \ref{fig:r_v423}, V423 Aur and HD34974. HD34974 is not 
a variable star. 
So by comparing with HD34974, we can know 
whether V423 Aur has photometric changes .
Figure \ref{fig:09gri} 
and Figure \ref{fig:11gri} show the results of 
differential photometry.  By comparing the light curves of $g$ band 
and $r$ band, the $g$ magnitudes and $r$ magnitudes do not show the same variation 
trend. So we do not think that we have observed the variability of V423 Aur. 
The larger magnitude scatter of $g$ and $r$ band in Figre \ref{fig:09gri} 
may be due to instrumentation or weather effects.
The light curves of $i$ band were not drawn because of its large error 
due to the instrument itself.

\begin{table}
    \centering
\caption{Observation of V423 Aur with NAGIOT}
\label{tab:2}      
\begin{tabular}{cccc}
\hline\noalign{\smallskip}
    Observation Date & Exposure & Filters & Observation times  \\
\noalign{\smallskip}\hline\noalign{\smallskip}
    2018-01-09 & 150s & $g,r,i$ & 29 \\
    2018-01-11 & 150s & $g,r,i$ & 27 \\
\noalign{\smallskip}\hline
\end{tabular}
\end{table}


\subsection{High Resolution Spectra (HRS) of V423 Aur}
\label{s:data:hrs}

At the same time of photometry, we also observed V423 Aur with high 
resolution spectrograph (R$\sim49800$) of 2.16m Telescope, which locates 
at Xinglong Observatory, China. 
The spectrograph covers the wavelength range from 4000$\mathrm{\AA} \sim 
10200 \mathrm{\AA}$. 
From 9th to 11st Jan. 2018, three nights, the high resolution 
spectra of H$\alpha$ were observed for 12 times, 11 times and 4 times. 

The spectral reductions were carried out with the IRAF 
package following the standard procedures: order identification, bias 
subtraction, flat-field correction, scattered-light subtraction, spectrum 
extraction, wavelength calibration (based on Th-Ar lamp spectra), and 
continuum normalization. 

\begin{figure*}                 
    \centering
    \includegraphics[width=\textwidth]{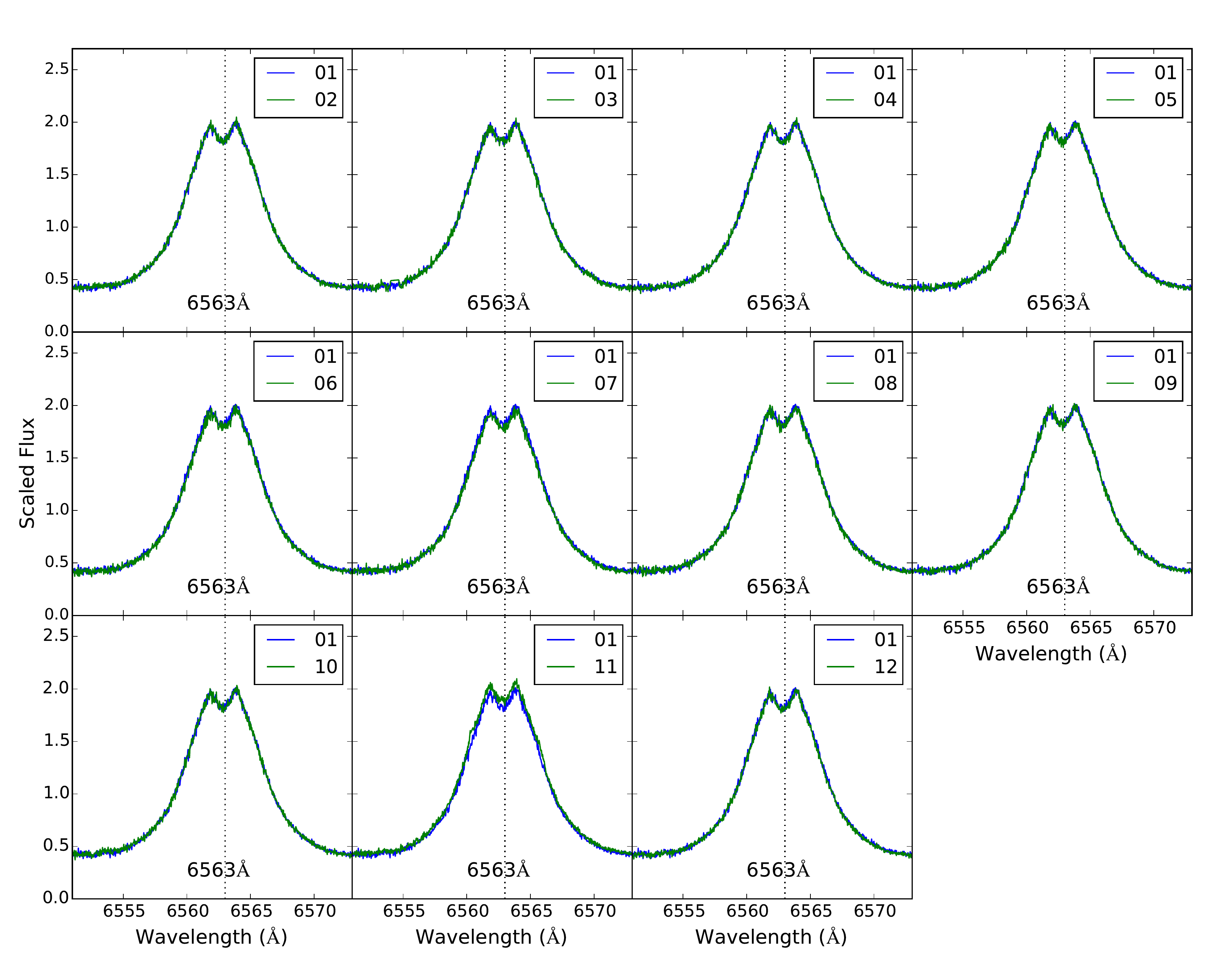}
    \caption{HRS of V423 Aur observed on 2018-01-09. 
    All other 11 spectra are compared to the first spectrum (blue line). 
    \label{fig:20180109_hrs}}
\end{figure*}

\begin{figure*}                                             
    \centering
    \includegraphics[width=\textwidth]{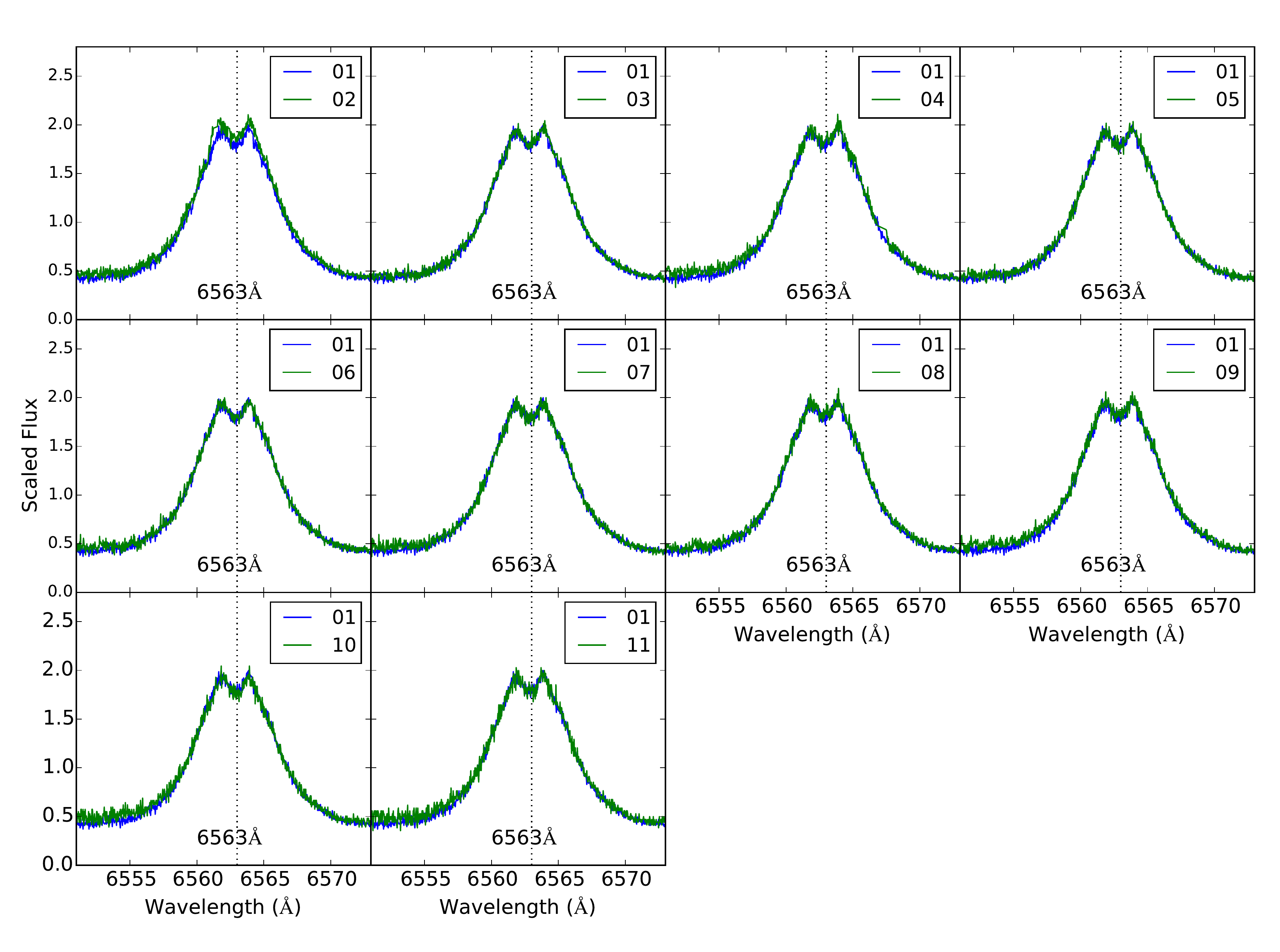}
    \caption{HRS of V423 Aur observed on 2018-01-10. 
    All other 10 spectra are compared to the first spectrum (blue line).
    \label{fig:20180110_hrs}}
\end{figure*}

\begin{figure*}                                                     
    \centering
    \includegraphics[width=\textwidth]{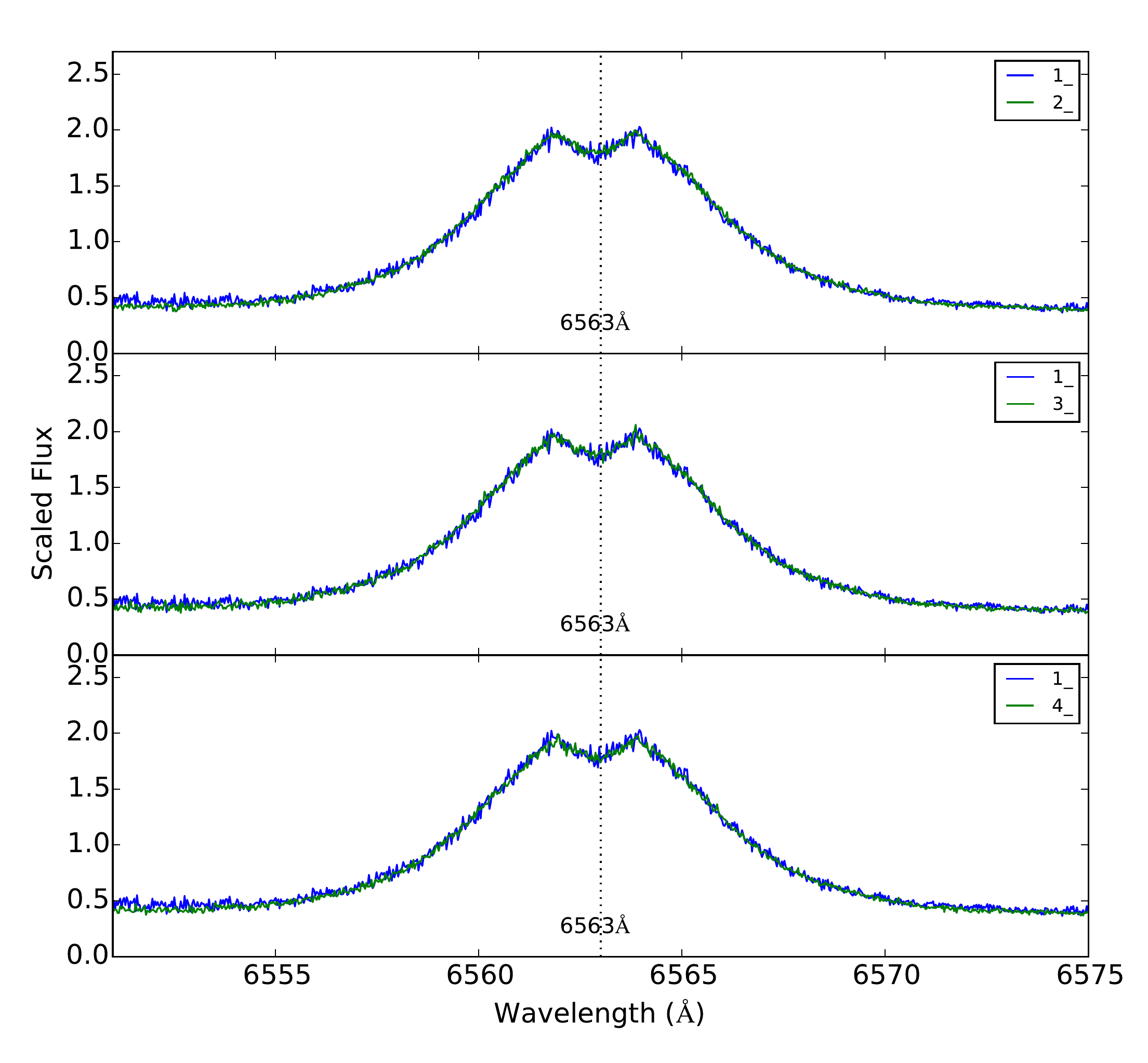}
    \caption{HRS of V423 Aur observed on 2018-01-11. 
    All other 3 spectra are compared to the first spectrum (blue line).
    \label{fig:20180111_hrs}}
\end{figure*}

In order to detect the variation of H$\alpha$ intensity, we compared 
all other 
11 spectra to the spectrum of first observation, see Figure 
\ref{fig:20180109_hrs}. We do the same operation for the spectra of 
10th Jan and 11th Jan (Figure \ref{fig:20180110_hrs}, 
Figure \ref{fig:20180111_hrs}). From Figure \ref{fig:20180109_hrs}, 
\ref{fig:20180110_hrs} and \ref{fig:20180111_hrs}, we can't detect the 
variation of H$\alpha$ intensity. 


\section{Results}
\label{s:results}

\begin{figure*}                                             
    \centering
    \includegraphics[width=1.2\textwidth]{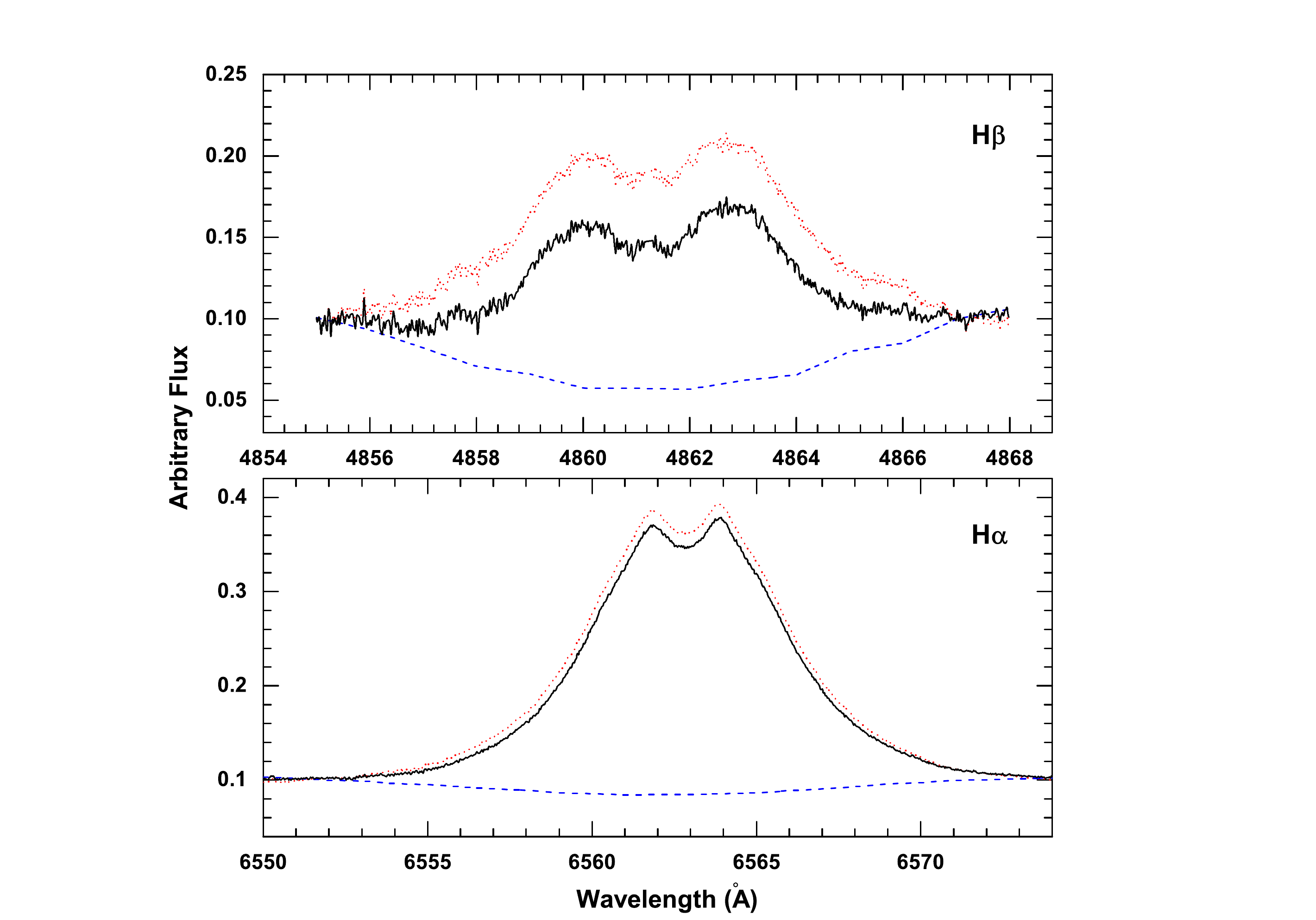}
    \caption{The H$\beta$ (upper panel) and H$\alpha$ (bottom panel) emission line profiles. 
    In each panel, the observed profiles and
the modeled absorptions are plotted as the black solid 
and blue dashed lines, respectively. The red dotted lines
represent the residual profiles after removing the modeled absorptions.
    \label{fig:HaHb}}
    \end{figure*}

\begin{figure*}
    \centering
    \includegraphics[width=0.8\textwidth]{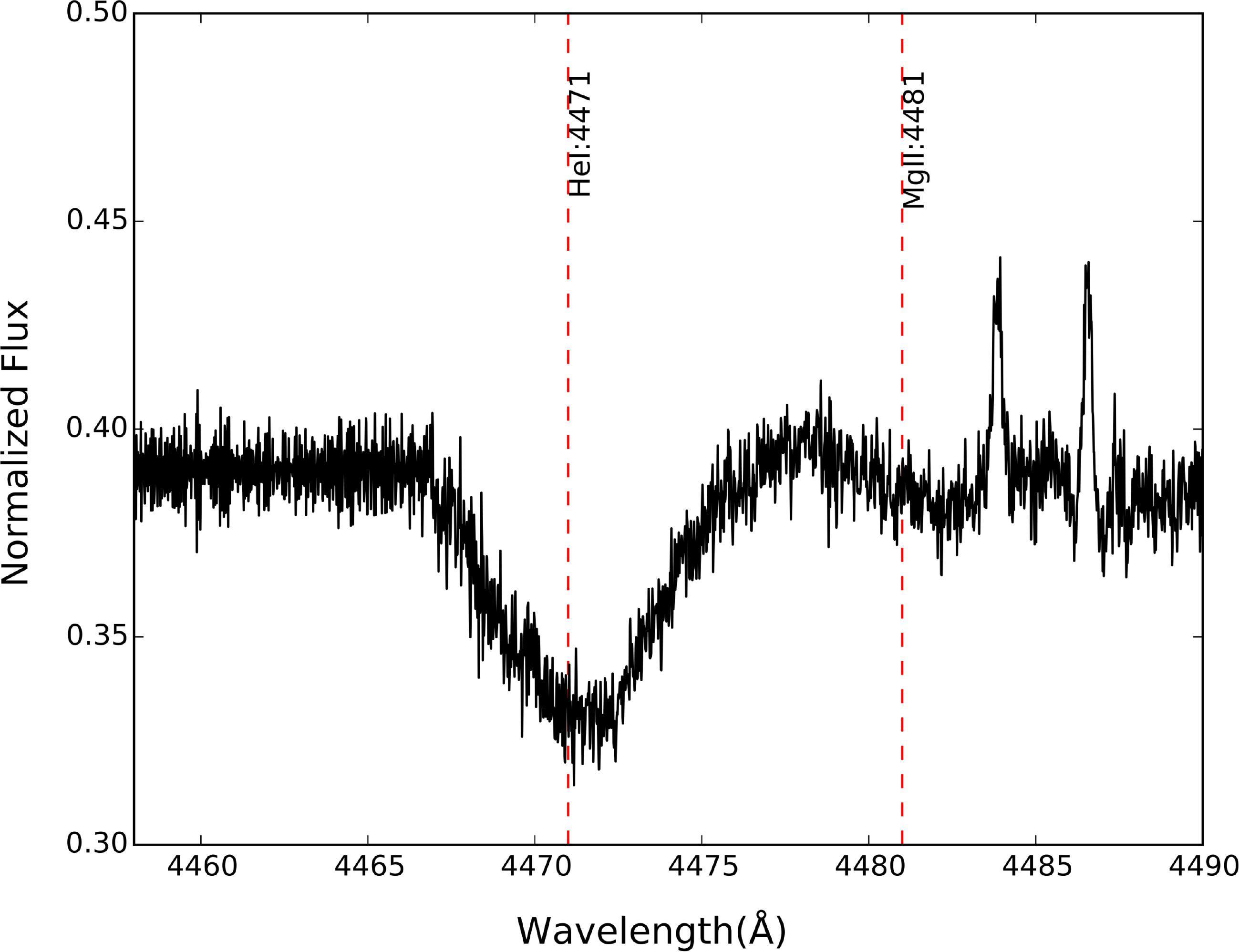}
    \caption{HRS of V423 Aur observed on 2018-01-10.  
    The marked HeI absorption line ($\lambda 4471 \mathrm{\AA}$) and MgII 
    absorption line ($\lambda 4481 \mathrm{\AA}$) can be used to classify star.
    \label{fig:heimgii}}
\end{figure*}

V423 Aur was detected to have a companion by \citet{2019Kervella}. 
\citet{1986halbedel} reported that the V magnitde of 
V423 Aur is 8.68.
A webpage (https://www.universeguide.com/star/aurigae) concluded 
that the star has a B-V Colour Index 
of 0.09 which means the star's temperature can be calculated 
at being about 9100 K. However, V423 Aur is a variable star, 
its B and V magnitude are not credible. 
We re-certified its spectral type based on HRS. 
Figure \ref{fig:heimgii} is segment of the HRS of V423 Aur. Its 
wavelength ranges from 4460$\mathrm{\AA}$ to 4490$\mathrm{\AA}$.  
HeI ($\lambda$4471) $ and $ MgII ($\lambda$4481) lines are within 
this wavelength range.  Figure \ref{fig:heimgii} shows 
obviously that HeI ($\lambda$4471) $>$ MgII ($\lambda$4481). 
By checking the HeI ($\lambda$4121) line, it is absent. So based on the 
criteria of classifying OB stars mentioned by 
\citet{2019Liu}, V423 Aur is determined as a "B7 V" star with 
temperature about 10300 Kelvin, 
which is a little different with the result (B8) from Simbad. 

The webpage above also show that the brightness of V423 Aur 
ranges from a magnitude of 8.791 to 8.633 over its variable period. 
Its variable/pulsating period lasts for 0.2 days. In our photometric 
observations, we did not find any change in brightness.
Using the newly released GAIA data,  the distance of this star is 
estimated to be 1442 pc \citep{2018Bailer}.

The H$\beta$ and H$\alpha$ line emissions corrected by a theoretical ATLAS9
model atmosphere with effective temperature of 
T$_{eff}$ =10,000 K and log {\it g} = 4 cm~s$^{-2}$ \citep{2002Heiter} described 
by absorption profiles are shown in Figure \ref{fig:HaHb}. The H$\alpha$/H$\beta$ flux 
ratio measured from V423 Aur is 5.61$\pm$0.19. By comparing the observed 
H$\alpha$/H$\beta$ intensity ratio with the theoretical value at T$_{e}$ =10$^{4}$ K 
and n$_{e}$ =10$^{4}$ cm$^{-3}$ \citep{1987Hummer} and by using the reddening 
law of \citet{1983Howarth} for R$_{V}$ = 3.1, we can derive the extinction coefficient 
E(B-V)= 0.709$\pm$0.036 for this object, which is consistent with the previously reported 
values of E(B-V)= 0.671 \citep{2019Kervella} estimated from local interstellar medium. This 
result indicates that the target is associated with regional nebulae.

\begin{figure*}                                                              
    \centering
    \includegraphics[width=0.8\textwidth]{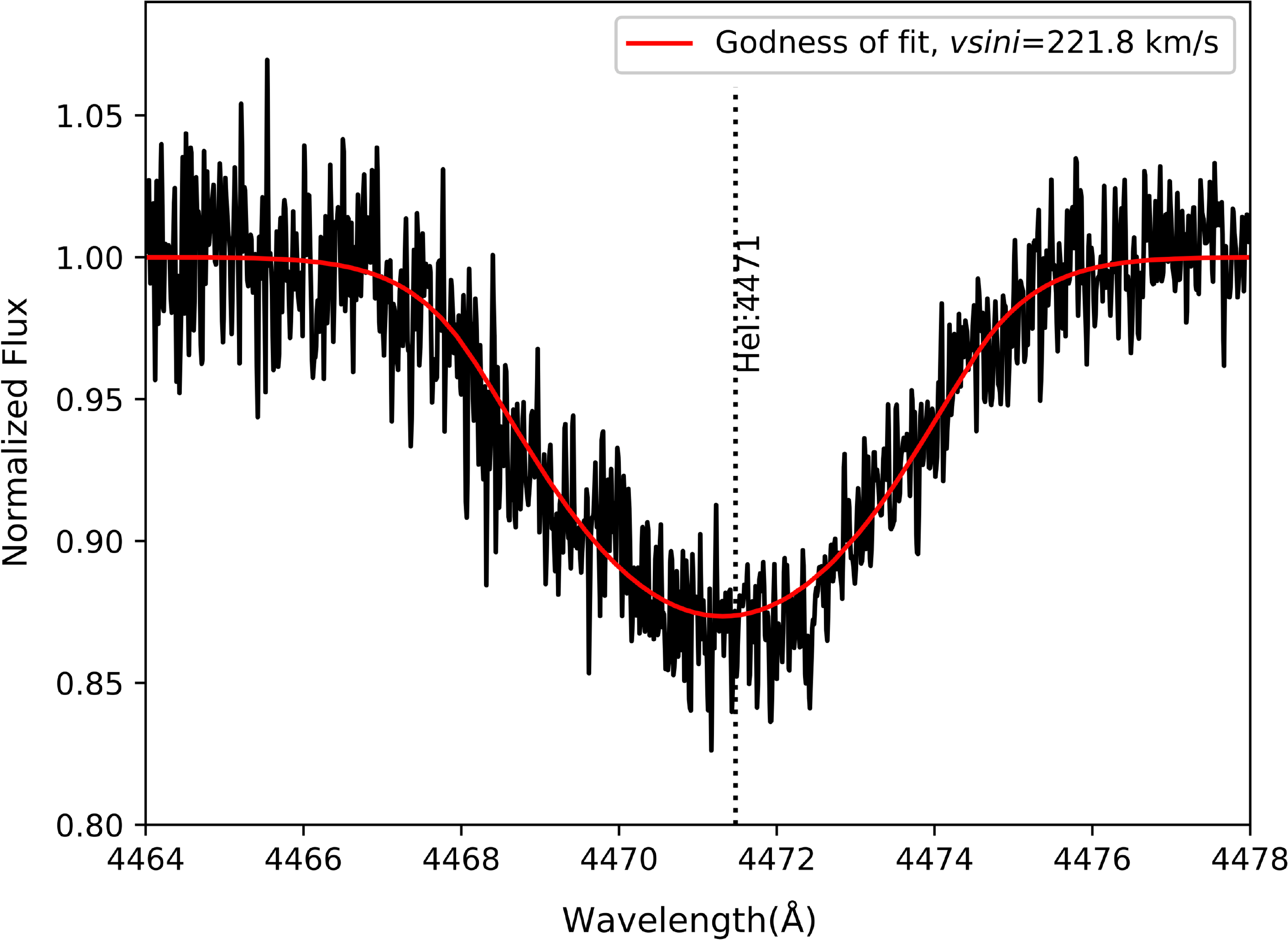}
    \caption{HRS HeI absorption line of V423 Aur. This line is 
    fitted to estimate the $v$sin$i$ of V423 Aur.
    \label{fig:HeI}}
\end{figure*}

By checking other emission 
and absorption lines of V423 Aur, we found the HeI absorption line centered 
on 6678.15$ \mathrm{\AA}$ was broadened (Figure \ref{fig:red}). The broadened absorption line can 
be explained by stellar rotation. To estimate the stellar rotation ($v\mathrm{sin}i$), 
then we fitted HeI ($\lambda 4471 \mathrm{\AA}$) absorption line which is from HRS by  
using a versatile and user-friendly IDL tool \citep{2014simon}. The tool is based on a combined Fourier transform 
and  goodness-of-fit methodology for the line-broadening characterization in OB-type stars. 
Here we did not use HeI ($\lambda 6678.15 \mathrm{\AA}$) to do fitting because it is mixed with FeI 
line at 6678.15$ \mathrm{\AA}$ (Figure \ref{fig:red}). 
 Fitting result shows that $v\mathrm{sin}i$ is about 221.8 km/s (Figure \ref{fig:HeI}), 
 which is consistent with the general rotation velocity of Be star \citep{1965McNally}. 

\section{Discussion}
\label{s:dis}
As described in the previous sections, the medium-resolution spectra of 
V423 Aur show irregular 
changes in H$\alpha$ intensity. However, no changes were detected in the 
follow-up photometry and HRS of V423 Aur. The photometry 
and HRS were observed on the same day, about one month after the 
observation of medium-resolution spectra. So, the changes in intensity may 
be due to 
transient phenomena of the star itself. That is to say, the transient 
phenomena was just recorded during MRS observation of V423 Aur. 

How to understand the transient phenomena?  The circumstellar disk or 
gas shell is very common for Be star\citep{2013Rivinius, 1995hofner}. 
The H$\alpha$ emission 
line is from surrounding excited gas. So we explain the irregular H$\alpha$ 
variation as follow. The change in the intensity 
of the H$\alpha$ emission line is due to the ejection of 
matter from the gas shell or disk. Random ejection of matter leads 
to irregularity of H$\alpha$ lines. The broader blue wings of H$\alpha$ lines 
observed by LAMOST (No. 4, No. 5, No. 6 and No. 7 of Figure \ref{fig:Ha}) 
mean that the matter was ejected toward us. The ejection process was a 
short-term event, which was just observed by LAMOST and has been 
finished when we observed it again with NAGIOT and 2.16m telescope. 
So we can not find any intensity changes from photometry and HRS.

\section{Summary}
\label{s:summary}
On Dec. 5th, 2017, in LAMOST Medium-Resolution test observation, a Be 
star (V423 Aur), which is located in the constellation of Auriga, showing 
irregular changes in H$\alpha$ intensity was observed. V423 Aur was 
observed for seven times by LAMOST. By comparing these 7 results, it 
shows different H$\alpha$ intensity. About one month later, we 
observed V423 Aur with NAGIOT and 2.16m telescope High Resolution 
spectrograph. Based on HRS, the Be star was determined as a B7V type 
star and associated with regional nebulae. 
However, there is no any changes in H$\alpha$ intensity from HRS of 2.16m 
telescope. We also didn't detect any magnitude variation in photometric results of NAGIOT. 
So we think the change in the H$\alpha$ intensity is a transient phenomenon.
The variable H$\alpha$ line may be due to the ejection of matter from 
gas shell or disk. The ejection process has finished in the follow-up 
observation about one month later. Since Sep. 2018, LAMOST started the 
second stage survey program (MRS). More medium resolution spectra of Be stars 
will be observed. We plan to repeat the observations of V423 Aur and other similar 
Be stars to check if these LPVs are real.

\normalem

\clearpage
\begin{acknowledgements}
    This project is supported by the National Key R \& D Program of China (No. 2017YFA0402704), 
    and the National Natural Science Foundation of China (grant Nos. 11733006, 11403061, 
    11403037, 11225316, 11173030, 11303038, Y613991N01 and U1531245). 
    the Key Laboratory of Optical 
    Astronomy, National Astronomical Observatories, Chinese 
    Academy of Sciences, Key Research Program of Frontier Sciences，
    CAS，Grant NO. QYZDY-SSW-SLH007.
    
    C.-H. Hsia acknowledges the supports from The Science and Technology Development Fund, 
    Macau SAR (file no. 119/2017/A3, 061/2017/A2, and 0007/2019/A) and Faculty Research Grants 
    of the Macau University of Science and Technology (program no. FGR-19-004-SSI). 
    
    Guoshoujing Telescope (the Large Sky Area Multi-Object Fiber Spectroscopic 
    Telescope LAMOST) is a National Major Scientific Project built by the 
    Chinese Academy of Sciences. Funding for the project has been provided 
    by the National Development and Reform Commission. LAMOST is operated 
    and managed by the National Astronomical Observatories, Chinese Academy 
    of Sciences.
    
    We acknowledge the support of the staff of the Xinglong 2.16m 
    and 1.26m telescope. This work was partially supported by 
    the Open Project Program of the Key Laboratory of Optical 
    Astronomy, National Astronomical Observatories, Chinese 
    Academy of Sciences.

\end{acknowledgements}

\bibliographystyle{raa}
\bibliography{ms0248}

\begin{thebibliography}{26}
\providecommand\natexlab[1]{#1}
\providecommand\JournalTitle[1]{#1}

\bibitem[{Baade}(1982)]{1982Baade}
{Baade}, D. 1982, \aap, 105, 65

\bibitem[{Bailer-Jones} {et~al.}(2018)]{2018Bailer}
{Bailer-Jones}, C.~A.~L., {Rybizki}, J., {Fouesneau}, M., {Mantelet}, G., \&
  {Andrae}, R. 2018, \aj, 156, 58

\bibitem[{Balona}(1975)]{1975Balona}
{Balona}, L.~A. 1975, \mnras, 173, 449

\bibitem[{Balona}(1990)]{1990Balona}
{Balona}, L.~A. 1990, \mnras, 245, 92

\bibitem[{Balona}(1995)]{1995Balona}
{Balona}, L.~A. 1995, \mnras, 277, 1547

\bibitem[{Balona} {et~al.}(1992)]{1992Balona}
{Balona}, L.~A., {Cuypers}, J., \& {Marang}, F. 1992, \aaps, 92, 533

\bibitem[{Catelan} \& {Smith}(2015)]{2015catelan}
{Catelan}, M., \& {Smith}, H.~A. 2015, {Pulsating Stars}


\bibitem[{Conroy} {et~al.}(2018)]{2018conroy}
{Conroy}, C., {Strader}, J., {van Dokkum}, P., {et~al.} 2018, arXiv:1804.05860

\bibitem[{Cuypers} {et~al.}(1989)]{1989Cuypers}
{Cuypers}, J., {Balona}, L.~A., \& {Marang}, F. 1989, \aaps, 81, 151

\bibitem[{Eyer} \& {Mowlavi}(2008)]{2008eyer}
{Eyer}, L., \& {Mowlavi}, N. 2008, in Journal of Physics Conference Series,
  Vol. 118, Journal of Physics Conference Series, 012010

\bibitem[{Halbedel}(1986)]{1986halbedel}
{Halbedel}, E.~M. 1986, Information Bulletin on Variable Stars, 2919

\bibitem[{Heiter} {et~al.}(2002)]{2002Heiter}
{Heiter}, U., {Kupka}, F., {van't Veer-Menneret}, C., {et~al.} 2002, \aap, 392,
  619

\bibitem[{Hoefner} {et~al.}(1995)]{1995hofner}
{Hoefner}, S., {Feuchtinger}, M.~U., \& {Dorfi}, E.~A. 1995, \aap, 297, 815

\bibitem[{Howarth}(1983)]{1983Howarth}
{Howarth}, I.~D. 1983, \mnras, 203, 301

\bibitem[{Hummer} \& {Storey}(1987)]{1987Hummer}
{Hummer}, D.~G., \& {Storey}, P.~J. 1987, \mnras, 224, 801

\bibitem[{Jaschek} {et~al.}(1981)]{1981Jaschek}
{Jaschek}, M., {Slettebak}, A., \& {Jaschek}, C. 1981, {Be star terminology.},
  Be Star Newsletter

\bibitem[{Kervella} {et~al.}(2019)]{2019Kervella}
{Kervella}, P., {Arenou}, F., {Mignard}, F., \& {Th{\'e}venin}, F. 2019, \aap,
  623, A72

\bibitem[{Liu} {et~al.}(2019)]{2019Liu}
{Liu}, Z., {Cui}, W., {Liu}, C., {et~al.} 2019, arXiv e-prints,
  arXiv:1902.07607

\bibitem[{Lucy}(1974)]{1974Lucy}
{Lucy}, L.~B. 1974, \aj, 79, 745

\bibitem[{McNally}(1965)]{1965McNally}
{McNally}, D. 1965, The Observatory, 85, 166

\bibitem[{Porter} \& {Rivinius}(2003)]{2003Porter}
{Porter}, J.~M., \& {Rivinius}, T. 2003, \pasp, 115, 1153

\bibitem[{Rivinius} {et~al.}(2013)]{2013Rivinius}
{Rivinius}, T., {Carciofi}, A.~C., \& {Martayan}, C. 2013, Astronomy and
  Astrophysics Review, 21, 69

\bibitem[{Sim{\'o}n-D{\'\i}az} \& {Herrero}(2014)]{2014simon}
{Sim{\'o}n-D{\'\i}az}, S., \& {Herrero}, A. 2014, Astronomy and Astrophysics,
  562, A135

\bibitem[{Smith}(2001)]{2001Smith}
{Smith}, M.~A. 2001, \apj, 562, 998

\bibitem[{Struve}(1931)]{1931Struve}
{Struve}, O. 1931, \apj, 73, 94

\end{thebibliography}

\end{document}